\documentclass[prl,reprint,showpacs]{revtex4-1}
\usepackage{graphicx}
\usepackage{amsmath}

\renewcommand{\vec}[1]{\mathbf{#1}}

\begin{document}
\title{Quantized conductance through the quantum evaporation of bosonic atoms}
\author{D.J. Papoular$^{1,2}$, L.P. Pitaevskii$^{1,3}$, and S. Stringari$^{1}$}
\affiliation{${}^{1}$INO-CNR BEC Center and Dipartimento di Fisica, 
  Universit\`a di Trento, 38123 Povo, Italy}
\affiliation{${}^{2}$LPTM, UMR 8089 of CNRS and Universit{\'e} 
  de Cergy--Pontoise, 95302 Cergy--Pontoise, France}
\affiliation{${}^{3}$Kapitza Institute for Physical Problems, Kosygina
2, 119334 Moscow, Russia}
\date{\today}


\begin{abstract}
We analyze theoretically the quantization of conductance
occurring with cold bosonic atoms trapped
in two reservoirs connected by a constriction with
an attractive gate potential.
We focus on temperatures slightly above
the condensation threshold in the reservoirs.
We show that a conductance step occurs,
coinciding with the appearance of a condensate in the
constriction.
Conductance relies on a collective process involving the
quantum condensation of an atom into an elementary excitation
and the subsequent quantum evaporation of an atom, in contrast with
ballistic fermion transport.
The value of the bosonic conductance plateau is strongly enhanced
compared to fermions and
explicitly 
depends on temperature.
We highlight the role of the repulsive interactions 
between the bosons in preventing
them from collapsing into the constriction.
\end{abstract}

\maketitle

In mesoscopic systems, where the motion of quantum particles occurs
over distances of the order of their coherence length, transport
phenomena exhibit quantum signatures \cite{nazarov:CUP2009}.
The quantization of conductance \cite{vanhouten:PhysToday1996}
is a hallmark among these effects. It
reflects the discrete nature of the transport
channels inside a strongly constricted geometry, and it occurs 
if the spread in energies of the 
incident particle distribution
is smaller than the energy separation of these channels. 
It was first observed in electronic transport through a quantum
point contact \cite{vanwes:PRL1988} as a series of plateaux in the 
conductance when the distance between the gate electrodes was increased.
In this fermionic case, the conductance quantum $G_K=e^2/h$
involves
fundamental constants only, which makes it
relevant for metrology \cite[chap.~7]{nawrocki:Springer2015}.
Unlike the quantum Hall effect \cite{avron:PhysicsToday2003}, 
it occurs in the absence of a magnetic field
and has been predicted
to affect neutral Helium atoms
\cite{sato:JLTP2005}. 

\begin{figure}
  \centering
  \includegraphics[width=.7\linewidth]{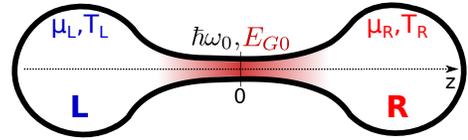}
  \caption{\label{fig:constriction}
    Two reservoirs ($L$, $R$) can exchange particles 
    through a smoothly tapered constriction inside which
    the spatially--dependent and attractive gate potential $E_G$ is varied.
  }
\end{figure}

Conductance quantization has recently been observed
in ultracold fermionic gases \cite{krinner:Nature2015}.
Atomic gases allow for a clean observation
in a simple setup involving two reservoirs connected by a constriction
within which an attractive gate potential
$E_G<0$ is varied
(see Fig.~\ref{fig:constriction}).
Experiments on ultracold fermions
aim at simulating electronic systems using
neutral particles 
\cite{brantut:Science2012,stadler:Nature2012,brantut:Science2013,
krinner:Nature2015}.
In the fermionic experiment of Ref.~\cite{krinner:Nature2015},
conductance quantization has been observed
at temperatures much lower than both the Fermi temperature $T_F$
and the confinement energy 
of the constriction, in analogy with the original results
on electronic transport \cite{vanwes:PRL1988} where only particles 
near the Fermi surface take part in transport phenomena.
This raises the question of whether conductance quantization 
also affects bosons. Previous observations in an optical setup
\cite{montie:Nature1991,vanhouten:Analogies1990}
and predictions with  cold 
matter waves \cite{thywissen:PRL1999}
have focused on systems where
all particles have the same incident energy, 
mimicking fermionic transport 
at the Fermi energy.
To our knowledge, the
specific role of bosonic statistics in quantized conductance situations has
not yet been investigated.
Cold atom setups allow for the exploration of
mesoscopic physics 
in situations where the
Bose distribution plays a key role 
\cite{papoular:PRL2012,papoular:PRL2014,lee:arXiv2015}.
They are also expected to exhibit the phenomenon of
quantum evaporation, whereby an elementary excitation of a
superfluid reaches
its surface and causes the evaporation of a single atom.
This phenomenon
had so far been studied 
experimentally \cite{johnston:PRL1966,hope:PRL1984}
and theoretically 
\cite{anderson:PhysLett1968,dalfovo:PRL1995}
in superfluid ${}^4\mathrm{He}$, 
and we consider it for the first time 
in the context of
superfluid atomic gases.

In this Letter, we show that conductance quantization occurs 
with bosonic atoms as well, and that 
the Bose statistics strongly enhances 
the value of the conductance step compared to fermions.
Unlike for fermions, 
this value explicitly depends on temperature, 
and the effect occurs with bosons up to temperatures higher than 
with fermions.
Furthermore, we show that the underlying transmission
mechanism is very different from the fermionic case and leads to
the occurrence of a single conductance plateau as the gate potential 
is varied, coinciding with the appearance of a condensate in
the constriction.
Transmission through the constriction relies
on quantum condensation followed by quantum evaporation:
an atom impinging on one end of the constriction 
excites a phonon in the condensate, which
travels through the constriction and causes the evaporation of a single
atom at its other end.
Hence, transport through the constriction involves a collective 
mechanism, as in Ref.~\cite{gutman:PRB2012}.
However, we focus on weakly--interacting Bose gases with
temperatures $T$ slightly above 
the critical temperature $T_B$ in the reservoirs,
so that these contain a thermal gas and no superflow occurs,
in contrast to Refs.~\cite{karpiuk:PRA2012,gutman:PRB2012,simpson:PRL2014}. 

The two reservoirs $L$ and $R$ of Fig.~\ref{fig:constriction}
can exchange particles via a 
constriction 
of length $l_C$ produced by the
potential $V_C(r,z)$. 
At its most stringent point $z=0$, we model it by
the radial harmonic trap 
$V(r,0)=m\omega_0^2r^2/2$.
We assume that the 
gate potential $E_G(z)<0$ 
also reaches its maximum value $|E_{G0}|$ at $z=0$.

\begin{figure}
  \includegraphics[angle=-90,width=.7\linewidth]
  {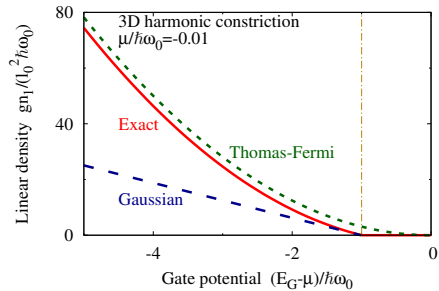}
  \caption{\label{fig:condprofile_harm}
  Linear condensate density at the center of the constriction
  as a function of the gate potential. The
  exact numerical result  (thick red line)
  interpolates in between the
  Gaussian approximation (dashed blue),  valid for 
  $|E_{G0}|\gtrsim\hbar\omega_0$, and the Thomas--Fermi result (dotted
  green), holding for large $|E_{G0}|$. 
}
\end{figure}
\emph{Equilibrium state.}
We first state two conditions on the strength of the 
interatomic interactions
which are
required for our analysis to hold for bosons.
These interactions should be 
\textit{(i)} weak enough for the reservoir thermodynamics to be dominated by
single--particle effects for temperatures $T\gtrsim T_B$, and
\textit{(ii)} strong enough to avoid a collapse of the system
into the attractive constriction.
These conditions are compatible and easily realized with
bosonic atoms trapped in box--like
potentials \cite{gaunt:PRL2013}.

\begin{figure*}
  \begin{minipage}{.32\textwidth}
    \includegraphics[angle=-90,width=\linewidth]{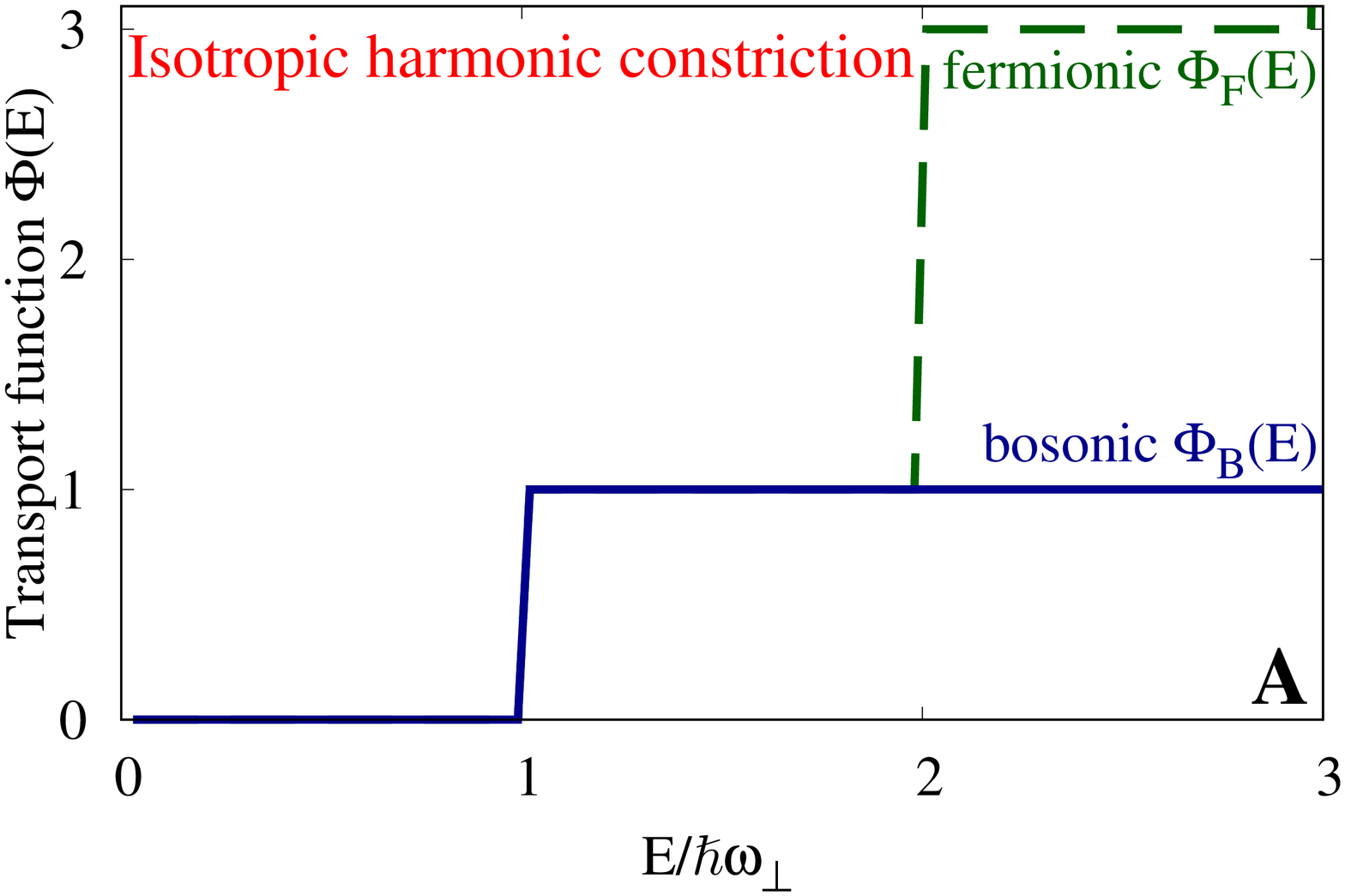}
  \end{minipage}
  \begin{minipage}{.32\textwidth}
    \includegraphics[angle=-90,width=\linewidth]{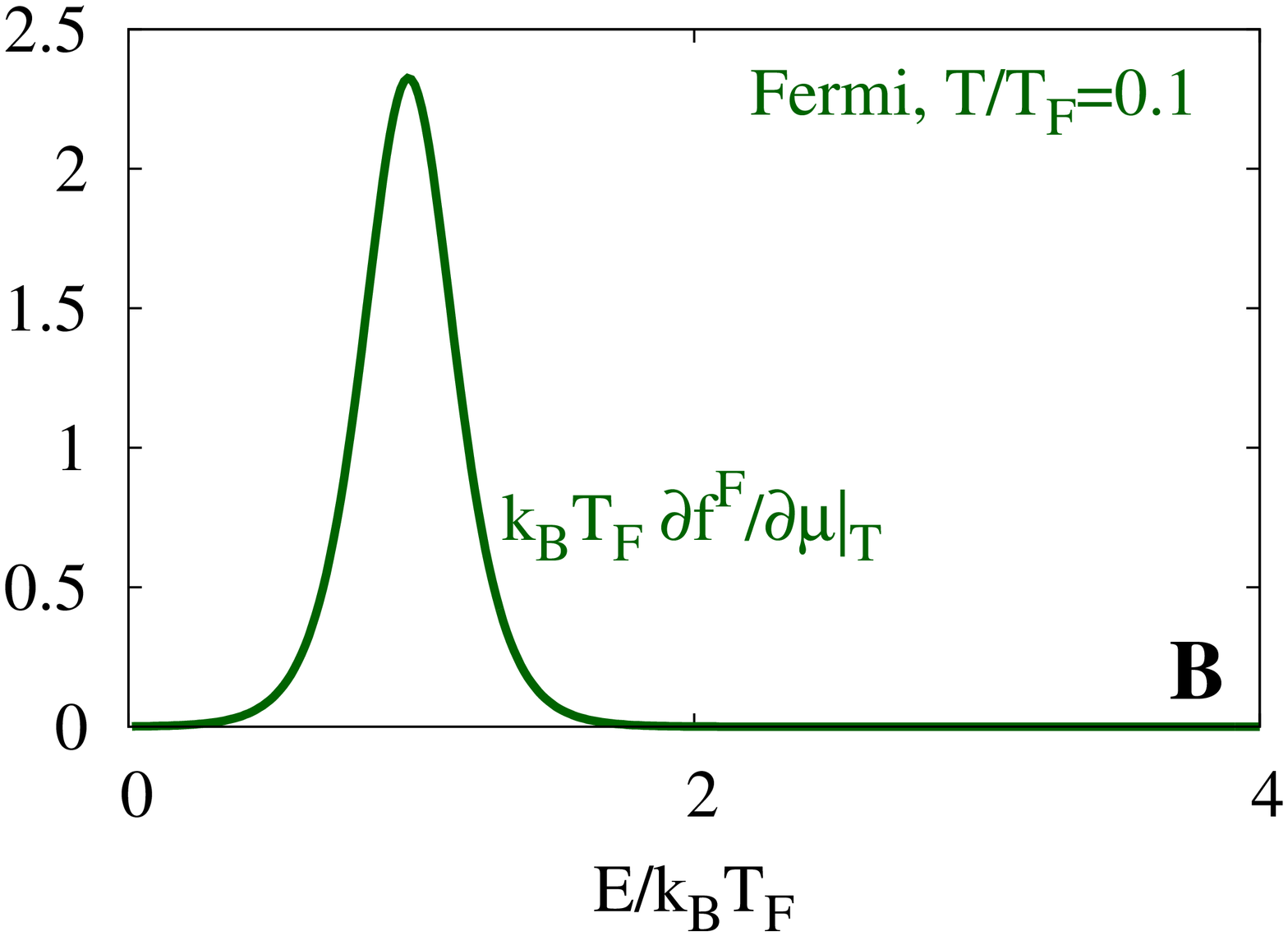}
  \end{minipage}
  \begin{minipage}{.32\textwidth}
    \includegraphics[angle=-90,width=\linewidth]{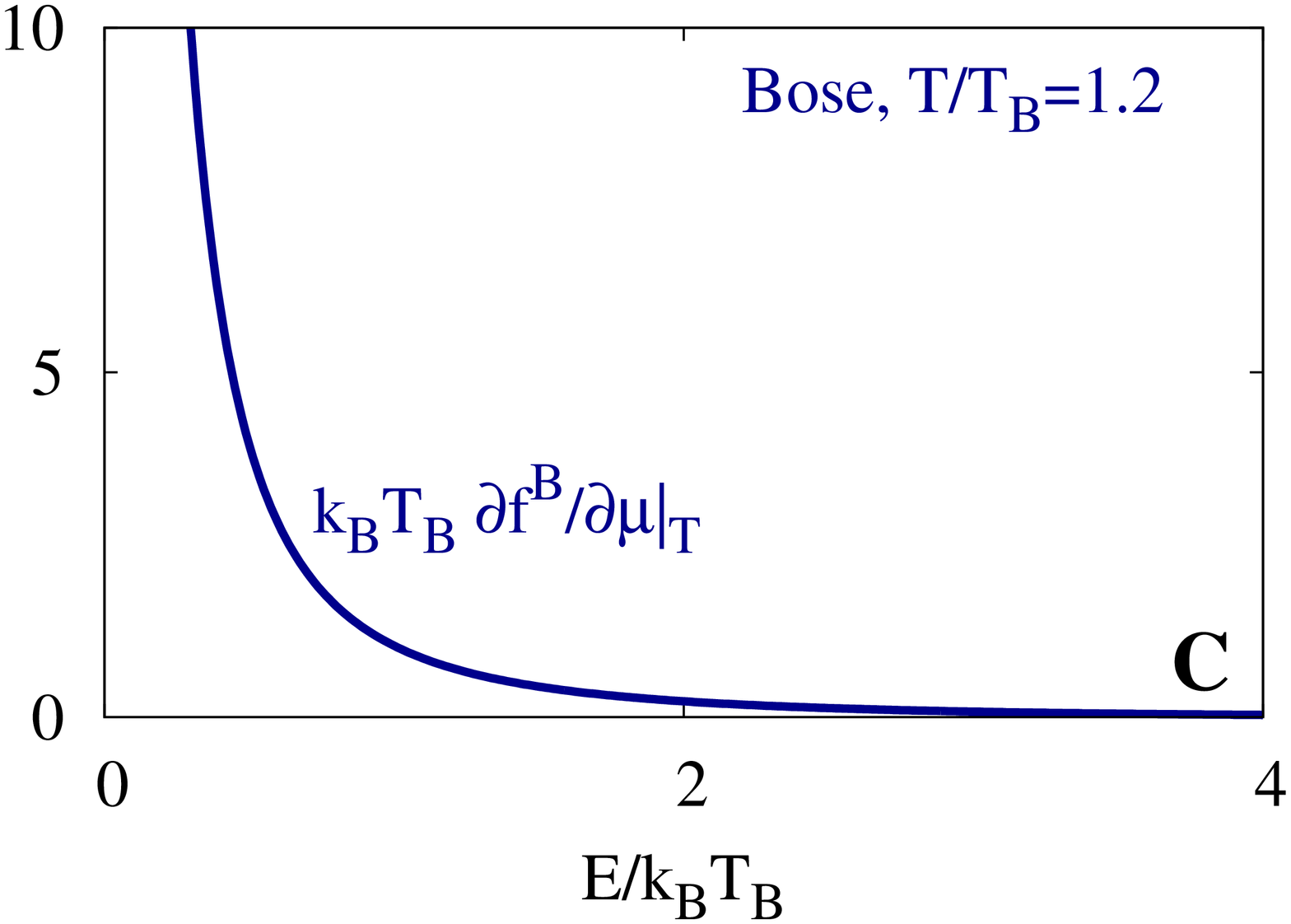}
  \end{minipage}
  \begin{minipage}{.32\textwidth}
  \end{minipage}
  \caption{\label{fig:Phi_df}
    A: Transport function of an isotropic harmonic constriction
    for fermions (thin dashed line) 
    and for bosons  (full solid curve).
    B and C: derivatives $\partial f^F/\partial\mu|_T$ and 
    $\partial f^B/\partial\mu|_T$ of the Fermi ($T/T_F=0.1$)
    and Bose ($T/T_B=1.2$) distributions.
  }
\end{figure*} 
\textit{(i)}
The effects of weak interactions in uncondensed Bose gases 
are well described by Hartree--Fock theory 
\cite[chap.~13]{pitaevskii:OUP2016}.
It predicts the chemical potential
$\mu(n,T)=\mu^{(0)}(n,T) +2gn$,
with $\mu^{(0)} < 0$ being the ideal--gas value, 
$n$ the 
density, and $g>0$ the interaction strength. 
In this theory,
the Bose distribution reads
$f^B(E)=1/[e^{(E+2gn-\mu)/k_\mathrm{B} T}-1]=1/[z^{-1}e^{E/k_\mathrm{B} T}-1]$, 
where the ideal--gas fugacity $z=\exp{(\mu^{(0)}/k_\mathrm{B}T)}$
and $E=p^2/2m$. The quantity  $\partial f^B/\partial\mu|_{T}$,
relevant for linear response,
can be replaced by
$\partial f^B/\partial\mu^{(0)}|_{T}$
if $2gn\kappa_T\ll N$. Here,
$N$ is the atom number in one reservoir,
and the isothermal compressibility
$\kappa_T=\partial N/\partial\mu|_T$ is
linked to its 
ideal--gas value by  
$N/\kappa_T= 2gn +N/\kappa_T^{(0)}$.
For $T\gtrsim T_B$,
$\kappa_T^{(0)} k_\mathrm{B}T_B/N= 
\sqrt{\pi}/[\zeta(3/2)\sqrt{1-z}]$, and the condition
$2gn \kappa_T\ll N$ means
$1-z \gg (gn/k_\mathrm{B}T_B)^2 4\pi/ \zeta^2(3/2)$.   
For a uniform gas, 
this condition is well satisfied for $T/T_B\geq 1.1$.
We focus on box--trap reservoirs which, for Bose gases, are more
favorable than the harmonically--trapped case, as interactions
play a weaker role within uniform gases ($gn/k_\mathrm{B}T_B\approx 0.02$)
than in trapped geometries 
($gn/k_\mathrm{B}T_B\approx 0.2$) \cite{papoular:PRL2014}.
Thus, we can describe the atoms in the reservoirs
as an ideal Bose gas with $\mu$ negative and small. 
We take $\mu/\hbar\omega_0\approx -0.01$ in the following.

\textit{(ii)}
Despite the assumption $T>T_B$, condensation 
occurs in the
constriction \cite{pinkse:PRL1997,stamperkurn:PRL1998,vandruten:PRL1997} 
if the gate potential $E_{G0}<-\hbar\omega_0+\mu$ 
is attractive enough for the energy 
of the first transverse state 
in the constriction to match the chemical potential 
of the gas in the reservoirs. Then,
in the absence of interactions,
the atoms would collapse into the constriction, impeding the investigation of
transport. 
The presence of weak repulsive
interactions between the bosons
prevents this collapse by making
the presence of too many
atoms in the constriction energetically disfavored. 
Neglecting the dilute thermal cloud, 
the condensate wavefunction $\Psi_0(r)$
at $z=0$, which depends only on the distance $r$ to the axis,
is the lowest--energy solution
to the Gross--Pitaevskii (GP) equation:
\begin{equation} \label{eq:GPE}
  (\mu-E_{G0})\Psi_0=
  (
    -\frac{\hbar^2}{2m}\Delta_r
    +\frac{1}{2}m\omega_0^2r^2
    +g|\Psi_0|^2
  )
  \Psi_0
  \ ,
\end{equation}
where the radial Laplacian satisfies 
$r\Delta_r\Psi_0=d(rd\Psi_0/dr)/dr$,
$g=4\pi\hbar^2 a/m$ and $a$ is
the scattering length encoding the interactions.
The density $|\Psi_0|^2$
at the point $z=0$ is determined by the effective chemical
potential $\mu-E_{G0}>0$. 
Figure \ref{fig:condprofile_harm} shows the linear density
$n_1=\int 2\pi r dr |\Psi_0|^2$ as a function of $E_{G0}$.
For $\mu-E_{G0}<\hbar\omega_0$, the constriction is empty.
For 
 $(\mu-E_{G0})$ just above $\hbar\omega_0$, 
the condensate wavefunction is nearly a Gaussian with the extent
$l_0=(\hbar/m\omega_0)^{1/2}$, and 
$gn_1/l_0^2=2\pi(\mu-E_{G0}-\hbar\omega_0)$. For more attractive gate
potentials, the Thomas--Fermi profile is quickly reached, leading to
$gn_1/l_0^2=\pi(\mu-E_{G0})^2/\hbar\omega_0$.
In all three cases, for $E_{G0}$ up to a few $\hbar\omega_0$,
the atom number 
in the constriction  $N_C < l_C/a$.
Hence, 
$N_C/N$ is small 
if the constriction is short enough, in which case
the atom number in the
reservoirs is unaffected by the small condensate in
the constriction.
On the other hand, the 1D density $n_1\gg a/l_0^2$, 
so that the condensate does not enter the
strongly--confined 1D regime \cite[chap.~24]{pitaevskii:OUP2016}.

\emph{Transport properties.}
We  focus on small deviations from the equilibrium situation where
both reservoirs are characterized by the same chemical potential
$\mu$ and temperature $T$. 
An important difference  between fermionic and bosonic transport
phenomena concerns the energies of the particles undergoing transport. 
In the
linear response regime, these are the energies 
for which the derivative $\partial f^{F,B}/\partial\mu|_{E,T}$ of the
(Fermi or Bose) distribution function 
with respect to $\mu$
is non--negligible. For fermions, this derivative is 
strongly peaked near the Fermi energy $k_\mathrm{B}T_F$ 
with a width $\sim k_\mathrm{B}T$
(see Fig.~\ref{fig:Phi_df}B), 
confirming the key role of the Fermi surface.
These fermions have non--vanishing energies and efficiently
traverse even sharply--defined constrictions \cite{szafer:PRL1989}.
By contrast,
for bosons, the derivative  $\partial f^{B}/\partial\mu|_{E,T}$
nearly
diverges for the energy $E=0$, and the mobile particles have energies
$\lesssim (1-z)k_\mathrm{B}T_B \approx |\mu|$ (see Fig.~\ref{fig:Phi_df}C).
This divergent behavior leads to the bosonic enhancement of conductance. 
Our choice of $T\gtrsim T_B$
means that $|\mu|\ll k_\mathrm{B}T_B$, and we assume in the following that
$k_\mathrm{B}T_B\lesssim\hbar\omega_0$,
hence, mobile bosonic atoms have energies $\ll \hbar\omega_0$.
Low--energy reflections at the ends of the constriction 
\cite[\S 25]{landau3:BH1977}
can be made negligible by smoothly connecting
it to the reservoirs \cite{glazman:JETP1988}
with a radius of curvature $R$ which is large 
compared to the characteristic atom wavelength
$(\hbar^2/m|\mu|)^{1/2}\sim 10l_0$.
Such a smoothly--tapered constriction 
was already used in the experiment of
Ref.~\cite{krinner:Nature2015}
where the Fermi momentum $k_F$ satisfies
$k_FR\sim 200$.

Introducing
the small differences in atom numbers, $\delta N=N_R-N_L$,
and chemical potentials, $\delta \mu=\mu_R-\mu_L$, between the reservoirs,
we define the isothermal conductance $G$ by the relation 
$\partial_t\delta N=-G\partial\delta\mu$.
The Landauer--B\"uttiker formalism 
\cite[chap.~1]{nazarov:CUP2009}
leads to the expression:
\begin{equation}
  \label{eq:conductance_isothermal}
  hG(E_G)=
  \int_0^{+\infty} dE 
  \,
  \Phi^{F,B}\left({E-E_G}\right)
  \,
  \left. \frac{\partial f^{F,B}}{\partial\mu}\right|_{E,T}
  \ .
\end{equation}
This equation 
holds for both fermionic  and bosonic  systems.
It is applicable whatever the reservoir geometry, 
encoded in the value of the degeneracy temperature $T_D=T_{F,B}$ 
\cite[chap.~10]{pitaevskii:OUP2016}. 
Equation (\ref{eq:conductance_isothermal}) shows that $G(E_G)$
is the convolution of two functions, which both depend on
the quantum statistics:
\textit{(i)} the transport function $\Phi^{F,B}(E)$ of the
constriction, and
\textit{(ii)} the derivative of the (Fermi or Bose) distribution function 
$f^{F,B}(E)=1/[z^{-1}\exp(E/k_\mathrm{B}T)\pm 1]$
of the reservoirs.

We first summarize the fermionic results 
of Ref.~\cite{krinner:Nature2015}.
Pauli exclusion ensures that the constriction remains
empty, so that transmission is a single--particle ballistic effect.
The transport function $\Phi^F(E)$, which
counts the  transport channels 
whose threshold energies are $\leq E$,
is determined by the most stringent part of the
constriction. It reads
$\Phi^F(E/\hbar\omega_0)=
\lfloor {E}/{\hbar\omega_0} \rfloor
(
  \lfloor {E}/{\hbar\omega_0}\rfloor +1
)/2$,
where $\lfloor x\rfloor$ stands for the integer part of $x$.
It exhibits jumps for energies that
are integer multiples $p\hbar\omega_0$ of the
constriction strength,  
reflecting the opening of additional transport channels
(dashed green line on Fig.~\ref{fig:Phi_df}A).
These jumps are the cause of the quantization of conductance.

We now consider bosonic atoms.
If the gate potential $E_{G0}>-\hbar\omega_0+\mu$, 
the constriction is empty (see Fig.~\ref{fig:condprofile_harm}). 
For sufficiently smooth spatial variations of 
$V_C(\vec{r})$ and $E_G(z)$, the motion of single thermal particles impinging
on it is quasiclassical \cite{glazman:JETP1988}. These experience a
repulsive barrier of height $(\hbar\omega_0+E_{G0})$, so that
low--energy
transmission through the constriction is blocked. 
Instead, for $E_{G0}<-\hbar\omega_0+\mu$,
the constriction is filled with a condensate
whose presence
strongly affects the nature of the
transport mechanism within the channel. 
The energies $\lesssim |\mu|$ of the incident atoms
are smaller than 
$gn_0$ at the center of the constriction, so that
transport is now a collective process.
It involves quantum condensation followed by quantum evaporation 
\cite{johnston:PRL1966,anderson:PhysLett1968,dalfovo:PRL1995},
which rely on the superfluidity of the condensate
and, hence, on the presence of interactions in between the bosons.
A thermal atom in a reservoir impinging on the constriction with
the energy $E$ condenses into
an elementary excitation inside the superfluid with the energy
$\epsilon=E-\mu$, which
crosses the constriction and evaporates an atom at its other end.
We describe this process using the Bogoliubov equations
\cite[chap.~12]{pitaevskii:OUP2016}: 
\begin{equation} \label{eq:bogoliubov}
  \begin{cases}
    \phantom{-}Eu &=
    (-\frac{\hbar^2}{2m}\Delta+V_\mathrm{ext}+2gn_0)u
    \phantom{-2\mu\,\,}
    +gn_0 v 
    \ ,
    \\
    -Ev &=
    (-\frac{\hbar^2}{2m}\Delta+V_\mathrm{ext}+2gn_0-2\mu)v
    +gn_0 u
    \ ,
    \\
  \end{cases}
\end{equation}
where $n_0(\vec{r})$ is the 3D condensate density,
the external potential
$V_\mathrm{ext}(\vec{r})=V_C(\vec{r})+E_G(z)$,
and $u(\vec{r}),v(\vec{r})$ are the 
Bogoliubov parameters. Equations~(\ref{eq:bogoliubov})
are valid for all values of $z$ 
and reduce to the Schr\"odinger
equation in the reservoirs where $n_0=0$. 
In the strongly constricted region near $z=0$, 
$n_0(r)$ is nearly the density profile of a condensate
trapped in the elongated radial harmonic trap 
$m\omega_0^2r^2/2$
with the effective chemical potential $(\mu-E_{G0})$.
Its excitation spectrum predicted by 
Eqs.~(\ref{eq:bogoliubov}) has multiple branches reflecting the 3D
geometry \cite{tozzo:NJP2003}. 
However, the condensate occupies 
the lowest--energy solution of the GP Eq.~(\ref{eq:GPE}), hence, its
low--energy excitations belong to the first
branch. 
For 
$|E_{G0}|/\hbar\omega_0\gtrsim 1.1$, the incident atoms have energies
$\lesssim |\mu| \ll gn_0$ and the excitations crossing  
the constriction are phononic.
Regardless of the value  of $E_{G0}$,
the second branch has the threshold energy
$2\omega_0$ 
\cite{tozzo:NJP2003,pitaevskii:PRA1997}, 
which is much greater than the incident energies,
so that this branch is never involved.
The smooth spatial variation of $V_\mathrm{ext}(\vec{r})$ ensures
a full conversion of the incident atoms into excitations of the
superfluid.
Transmission through the condensate--filled constriction 
hinges on this collective phenomenon: 
if the constriction is empty because the energy of its first
transverse mode is $>\mu$, bosonic transmission is blocked;
instead, if $E_G$ is sufficiently attractive for
the constriction to contain a condensate,
transmission is allowed and relies on
quantum condensation and evaporation. These two 
mechanisms lead
to a bosonic transport function which exhibits a single step:
$\Phi^B(E/\hbar\omega_0)=\Theta(E/\hbar\omega_0-1)$
(see Fig.~\ref{fig:Phi_df}A).

\begin{figure}
  \centering
  \includegraphics[angle=-90,width=.8\linewidth]{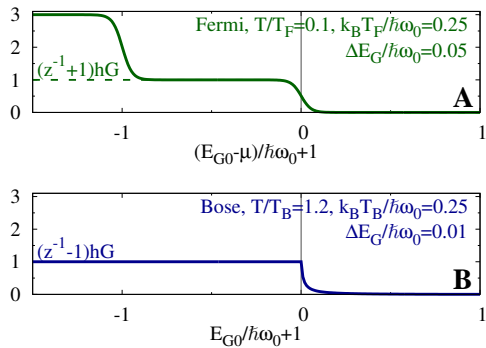}
  \caption{\label{fig:Phi_quantcondFB}
  Quantized conductance for 
  ultracold fermions (A, $T/T_F=0.1$)
  and cold bosons (B, $T/T_B=1.2$).
  In both cases, $\hbar\omega_0/k_\mathrm{B}T_D=4$.
  For fermions,
  the thick solid line is the exact solution $G(E_{G0})$
  and the thin dashed curve is the single--transport--channel
  prediction of Eq.~(\ref{eq:G1ch}). The results have been vertically
  rescaled by the step heights $1/(z^{-1}\pm 1)$.}
\end{figure}
\emph{Quantized conductance.}
The conductance $G(E_G/\hbar\omega_0)$
calculated from Eq.~(\ref{eq:conductance_isothermal}) 
depends on
$T/T_D$ and $\hbar\omega_0/k_\mathrm{B} T_D$.
We compare the fermionic and bosonic predictions on
Fig.~\ref{fig:Phi_quantcondFB} 
($T/T_F=0.1$ for fermions and $T/T_B=1.2$ for bosons;
$\hbar\omega_0/k_B T_D=4$ in both cases).
The fermionic prediction has the multiple step structure
observed in Refs.~\cite{vanwes:PRL1988,krinner:Nature2015}
due to the stepwise structure of the ballistic
$\Phi^F(E)$. 
By contrast, the bosonic graph exhibits one single step,
relating to the single step of $\Phi^B(E)$.
It occurs for $E_G=-\hbar\omega_0$ and, hence,
coincides with the appearance of the condensate in the constriction
(see Fig.~\ref{fig:condprofile_harm}).
For bosons, Eq.~(\ref{eq:conductance_isothermal}) can be integrated 
analytically;
an analogous results is obtained for the first fermionic step
by accounting for a single transport channel. We find:
\begin{equation}
  \label{eq:G1ch}
  hG^{F,B}=
  \begin{cases}
    \frac{1}
    {z^{-1}\exp[(E_G+\hbar\omega_0)/k_\mathrm{B}T]\pm 1}
    & \quad \text{if $E_G>-\hbar\omega_0$,}\\
    \frac{1}
    {z^{-1}\pm 1}
    &\quad \text{if $E_G\leq -\hbar\omega_0$,}
  \end{cases}
\end{equation}
where the $+$ and $-$ signs respectively apply to fermions
and bosons.
Equation (\ref{eq:G1ch})
reveals three differences between fermions and bosons,
concerning the step positions, their heights,
and the widths of the transition regions 
between two plateaux.
\textit{(i)}~For fermions,
the step is centered on $E_G=-\hbar\omega_0+\mu$, 
reflecting the key 
role of the Fermi surface at energies $\sim\mu$.
For bosons, 
the low--energy divergence discussed above causes the
step to occur at $E_G=-\hbar\omega_0$.  
\textit{(ii)}~For ultracold fermions, 
the fugacity $z\rightarrow\infty$, leading to the step height 
$1/(z^{-1}+1)\approx 1$.
Instead,
for bosons, $z\lesssim 1$ for $T\gtrsim T_B$,
leading to the very large step height 
$1/(z^{-1}-1)\approx 27$ for $T/T_B=1.2$.
\textit{(iii)}~For fermions, 
the 
width 
of the transition region is
$\Delta E_G^F\sim 2k_\mathrm{B}T$,
whereas the corresponding width for bosons is
$\Delta E_G^B\sim (1-z)k_\mathrm{B}T\approx|\mu|$. 
The conductance step is well defined if 
$\Delta E_G\ll \hbar\omega_0$.
Hence, Bose systems are greatly 
favored, as seen on Fig.~\ref{fig:Phi_quantcondFB} where
$k_BT/\hbar\omega_0$ is ten times as large for bosons than
for fermions, but the bosonic step width is quenched by the factor
$(1-z)$.

The conductance $G$ is positive, hence, the current 
$\partial_t\delta N$ opposes
the atom number difference $\delta N$,
which relaxes to equilibrium as 
$\delta N=\delta N_0 \exp(-t/\tau_1)$.
The decay time $\tau_1=\kappa_T/G$
is proportional to $N$ and is conveniently expressed in units of
$\tau_D=Nh/k_BT_D$.
Its measurement 
allows for an access to
$G(E_G/\hbar\omega_\perp)$.
It has recently been
measured with fermions \cite{krinner:Nature2015},
where $\kappa_T k_\mathrm{B}T_F/N=3/2$ at small $T$, so that 
$\tau_1=3\tau_D/2\sim$ a few seconds for the first conductance plateau.
For bosons, the isothermal compressibility diverges 
as one approaches the critical temperature,
but the stronger divergence of $G$ leads to shorter decay times 
$\tau_1=\tau_D(1-z)^{1/2}\sqrt{\pi}/\zeta(3/2)\sim$ a few hundred ms
for the single conductance plateau.

The quantization of bosonic conductance involving quantum evaporation
precludes its interpretation 
as the diffraction of atomic matter waves, in contrast with previous
studies
\cite{vanwes:PRL1988,montie:Nature1991,thywissen:PRL1999}. 
It also requires an attractive gate potential, 
unlike for fermions where 
conductance may be scanned by varying the constriction width
\cite{vanwes:PRL1988,krinner:Nature2015}.

The bosonic enhancement of conductance near the BEC transition
is the transport analogue of the enhancement of the isothermal 
compressibility. It is due to the possibility of 
accommodating multiple bosons in
the lowest--energy
transport channel, 
which is more populated
at temperatures closer to $T_B$. 
This enhancement signals a departure from
the fermionic conductance quantum $G_K=1/h$ observed both with electrons 
\cite{vanwes:PRL1988} and with neutral fermions \cite{krinner:Nature2015}.
Its observation in a regime where conductance is not quantized
has recently been reported \cite{lee:arXiv2015}.
Both the compressibility $\kappa_T$ and the conductance $G$, 
which diverge in the ideal--gas model, 
depend
on many--body effects in the critical region
near the transition \cite{verney:EPL2015}, 
where their characterization remains an open problem
both from the theoretical and experimental points of view. 
The measurement of the relaxation time $\tau_1$ in bosonic
systems with temperatures 
very close to
$T_B$ will provide more insight into these two quantities. 

Challenging open 
questions include 
\textit{(i)}
thermoelectric effects occurring through quantum evaporation,
\textit{(ii)}
conductance quantization in 2D bosonic
systems, where the quasicondensate
enhances the role of interactions 
\cite{prokofev:PRA2002,fletcher:PRL2015,desbuquois:PRL2014}, and
\textit{(iii)}
its impact in the presence of a superfluid,
whose investigation has been initiated by recent
experiments with strongly--interacting Fermi gases
\cite{stadler:Nature2012,husmann:arXiv2015,krinner:arXiv2015}.

\begin{acknowledgments}
We thank J.--P.~Brantut and T.~Esslinger for 
stimulating discussions. This work has been
supported by the European Research Council (ERC) through the QGBE grant,
by the QUIC grant of the Horizon 2020 FET program, and by
Provincia Autonoma di Trento.
\end{acknowledgments}

%\bibliography{quantcondbib}
%

\end{document}